\documentclass[11pt]{article}
\usepackage{amssymb,amsmath,latexsym}
\usepackage{mathrsfs}
\usepackage{tipa}
\usepackage{amsfonts}
\usepackage{graphicx}
\usepackage{subfigure}
\usepackage[title]{appendix}
\usepackage{color}
\usepackage{multirow}
\usepackage{amsbsy}
\usepackage{longtable}
\usepackage{floatrow}
\usepackage{indentfirst}
\usepackage{cite}
\usepackage{paralist}
\allowdisplaybreaks[4] 
\newtheorem{theorem}{Theorem}
\newtheorem{definition}{Definition}
\newtheorem{lemma}{Lemma}
\newtheorem{cor}{Corollary}
\usepackage[font=small]{caption}

\begin{document}
	\title{\bf Stable and robust $\ell_p$-constrained compressive sensing recovery via robust width property}
	\author {Zhiyong Zhou\footnote{Corresponding author, zhiyong.zhou@umu.se. This work is supported by the Swedish Research Council grant (Reg.No. 340-2013-5342). Mathematics Subject Classification (2010): 94A12, 94A20.}, Jun Yu\\
		Department of Mathematics and Mathematical Statistics, Ume{\aa} University, \\Ume{\aa},
		901 87, Sweden}
	\maketitle
	\date{}
	\noindent
$\mathbf{Abstract}$: We study the recovery results of $\ell_p$-constrained compressive sensing (CS) with $p\geq 1$ via robust width property and determine conditions on the number of measurements for standard Gaussian matrices under which the property holds with high probability. Our paper extends the existing results in Cahill and Mixon (2014) from $\ell_2$-constrained CS to $\ell_p$-constrained case with $p\geq 1$ and complements the recovery analysis for robust CS with $\ell_p$ loss function. \\
$\mathbf{Keywords}$: Compressive sensing; Robust width property; Robust null space property; Restricted isometry property.

\section{Introduction}
There has been enormous research interest in the field of compressive sensing (CS) since its advent in \cite{crt,d}. CS aims to recover an unknown signal with underdetermined linear measurements (see \cite{ek,fr} for a comprehensive view). Specifically, let $x^{\natural}$ be some unknown signal, which lies in a finite-dimensional Hilbert space $\mathcal{H}$, and let $\Phi:\mathcal{H}\rightarrow \mathbb{F}^m$ be the underdetermined measurement linear operator, where $\mathbb{F}$ is either $\mathbb{R}$ or $\mathbb{C}$. We denote the usual $\ell_p$ norm in 
$\mathbb{F}^m$ as $\lVert u \rVert_{\ell_p}=(\sum_{i=1}^m|u_i|^p)^{1/p}$ for $1\leq p<\infty$ and $\lVert u\rVert_{\infty}=\max_{1\leq i\leq m}|u_i|$. Then, given the measurements $y=\Phi x^{\natural}+e$ for some unknown error $e\in\mathbb{F}^m$ with $\lVert e\rVert_{\ell_2}\leq \varepsilon$ and $x^{\natural}$ is close to a particular subset $\mathcal{A}\subseteq\mathcal{H}$ that consists of members with some specific structure, there is a norm $\lVert\cdot\rVert_{\sharp}$ corresponding to $\mathcal{A}$ such that \begin{align}
\triangle_{\sharp,2}(y):=\arg\min\limits_{x}\,\,\lVert x\rVert_{\sharp}\,\,\,\text{subject to}\,\,\,\lVert\Phi x-y\rVert_{\ell_2}\leq \varepsilon
\end{align} 
is a good estimate of the original signal $x^{\natural}$, provided that the measurement matrix $\Phi$ satisfies certain properties (e.g., robust null space property, restricted isometry property). However, these conditions are not known to be necessary. Recently, for a broad class of triples $(\mathcal{H},\mathcal{A},\lVert\cdot\rVert_{\sharp})$, \cite{cm} completely characterized $\Phi$ for which \begin{align}
\lVert \triangle_{\sharp,2}(y)-x^{\natural}\rVert_2\leq C_0\lVert x^{\natural}-a\rVert_{\sharp}+C_1\varepsilon\,\,\,\forall\,\,x^{\natural}\in\mathcal{H}, e\in\mathbb{F}^m, \lVert e\rVert_{\ell_2}\leq \varepsilon, a\in\mathcal{A}
\end{align}
holds for any given constants $C_0$ and $C_1$ via robust width property (RWP). Here the $\lVert\cdot\rVert_2$ on the left hand side above denotes the norm induced by the inner product over $\mathcal{H}$. 

But in some certain cases, it is of special interest to use the $\ell_p$-norms with $1\leq p\leq \infty$ which is different from 2, as the loss function for measuring the noise level (see \cite{agr,dlr,jhf,sac,wllqy}). Hereafter, we will focus on the solutions of the corresponding $\ell_p$-constrained CS problem: \begin{align}
\triangle_{\sharp,p}(y):=\arg\min\limits_{x}\,\,\lVert x\rVert_{\sharp}\,\,\,\text{subject to}\,\,\,\lVert\Phi x-y\rVert_{\ell_p}\leq \varepsilon.
\end{align}
The cases that taking $\mathcal{H}$ to be either $\mathbb{R}^N$ or $\mathbb{C}^N$ and $\lVert\cdot\rVert_{\sharp}=\lVert \cdot\rVert_{\ell_1}$, and that taking $\mathcal{H}$ to be the Hilbert space of real (or complex) matrices with inner product $\langle X, Y\rangle=Tr(XY^{*})$ and considering the nuclear norm $\lVert\cdot\rVert_{\sharp}=\lVert \cdot\rVert_{*}$ have been studied via the corresponding robust null space properties in \cite{dlr} and \cite{kkrt}, respectively. In this paper, we complement these recovery analysis results via RWP, a kind of weaker property compared to the robust null space property (NSP) and restricted isometry property (RIP). And from the other point of view, we simultaneously extend the existing results of RWP for $\ell_2$-constrained CS in \cite{cm} to a much more general $\ell_p$-constrained case with $p\geq 1$. 

The remainder of the paper is organized as follows. In Section 2, we introduce the RWP and give the equivalent condition for stable and robust recovery results. We discuss the relationship between RWP and robust NSP in Section 3, and the relationship between RWP and RIP in Section 4. In Section 5, we obtain the optimal minimum required number of measurements for standard Gaussian matrices via RWP. We finish in Section 6 with a conclusion.  
\section{Robust width property}
Firstly, we present the definition of RWP which acts as our basic recovery analysis tool. As will be given below, our newly defined RWP involves a new parameter $p$ and it is obvious that the original RWP in \cite{cm} is a special case of ours with $p=2$.

\begin{definition}
 For any $p\geq 1$,	we say a linear operator $\Phi:\mathcal{H}\rightarrow \mathbb{F}^m$ satisfies the $(p,\rho,\alpha)$-RWP over the unit $\sharp$-ball $B_{\sharp}$ if $$
	\lVert x\rVert_2\leq \rho\lVert x\rVert_{\sharp}
	$$
	for every $x\in\mathcal{H}$ such that $\lVert\Phi x\rVert_{\ell_p}<\alpha\lVert x\rVert_2$. 
\end{definition}

In order to apply the results to various instances of CS, we adopt the definition of CS spaces from \cite{cm} here. 

\begin{definition}(Definition 1 in \cite{cm})
	A CS space $(\mathcal{H},\mathcal{A},\lVert\cdot\rVert_{\sharp})$ with bound $L$ consists of a finite-dimensional Hilbert space $\mathcal{H}$, a subset $\mathcal{A}\subseteq\mathcal{H}$, and a norm $\lVert\cdot\rVert_{\sharp}$ with the following properties: \\
	(i) $0\in\mathcal{A}$.\\
	(ii) For every $a\in\mathcal{A}$ and $v\in\mathcal{H}$, there exists a decomposition $v=v_1+v_2$ such that $$
	\lVert a+v_1\rVert_{\sharp}=\lVert a\rVert_{\sharp}+\lVert v_1\rVert_{\sharp},\,\,\lVert v_2\rVert_{\sharp}\leq L\lVert v\rVert_2.
	$$
\end{definition}

\noindent
{\bf Remark 1} As presented in Section 4 of \cite{cm}, a variety of examples of CS framework satisfy this framework, including weighted sparsity \cite{fmsy,ms,rw}, block sparsity \cite{de,ekb}, gradient sparsity \cite{nw1,nw2} and low-rank matrices \cite{kkrt}. \\

Next, we give the main result that completely characterizes the measurement operators $\Phi$ for which the stable and robust recovery results for $\ell_p$-constrained CS are achieved.

\begin{theorem}
	For any $p\geq 1$, any CS space $(\mathcal{H},\mathcal{A},\lVert\cdot\rVert_{\sharp})$ with bound $L$ and any linear operator $\Phi:\mathcal{H}\rightarrow \mathbb{F}^m$, the following statements are equivalent up to constants: \\
	(a) $\Phi$ satisfies the $(p,\rho,\alpha)$-RWP over $B_{\sharp}$.\\
	(b) For every $x^{\natural}\in\mathcal{H}$, $y=\Phi x^{\natural}+e$ with $e\in\mathbb{F}^m$ and $\lVert e\rVert_{\ell_p}\leq \varepsilon$, any solution $x^{\star}$ to  \begin{align}
    \min\limits_{x}\,\,\lVert x\rVert_{\sharp}\,\,\,\text{subject to}\,\,\,\lVert\Phi x-y\rVert_{\ell_p}\leq \varepsilon \label{min}
	\end{align} 
	satisfies $\lVert x^{\star}-x^{\natural}\rVert_2\leq C_0\lVert x^{\natural}-a\rVert_{\sharp}+C_1\varepsilon$ for every $a\in\mathcal{A}$. 
	
	\noindent
	In particular, (a) implies (b) with $C_0=4\rho$ and $C_1=\frac{2}{\alpha}$ provided $\rho\leq 1/(4L)$. Also, (b) implies (a) with $\rho=2C_0$ and 
	$\alpha=\frac{1}{2C_1}$.
\end{theorem}
 
 When taking $p=2$, this theorem goes to Theorem 3 in \cite{cm}. Its proof follows from Theorem 3 in \cite{cm} with minor modification. For the sake of completeness, we present the proof here. \\
 
 \noindent
 {\bf Proof.} $(a)\Rightarrow(b)$: Pick $a\in\mathcal{A}$ and decompose $v=x^{\star}-x^{\natural}=v_1+v_2$ according to (ii) in Definition 2 such that $\lVert a+v_1\rVert_{\sharp}=\lVert a\rVert_{\sharp}+\lVert v_1\rVert_{\sharp}$ and $\lVert v_2\rVert_{\sharp}\leq L\lVert v\rVert_2$. Then \begin{align*}
 \lVert a\rVert_{\sharp}+ \lVert x^{\natural}-a\rVert_{\sharp}&\geq \lVert x^{\natural}\rVert_{\sharp}\\
 &\geq \lVert x^{\star}\rVert_{\sharp} \\
 &= \lVert x^{\natural}+(x^{\star}-x^{\natural})\rVert_{\sharp}\\
 &= \lVert a+(x^{\natural}-a)+v_1+v_2\rVert_{\sharp}\\
 &\geq \lVert a+v_1\rVert_{\sharp}-\lVert(x^{\natural}-a)+v_2\rVert_{\sharp}\\
 &\geq \lVert a+v_1\rVert_{\sharp}-\lVert x^{\natural}-a\rVert_{\sharp}-\lVert v_2\rVert_{\sharp}\\
 &=\lVert a\rVert_{\sharp}+\lVert v_1\rVert_{\sharp}-\lVert x^{\natural}-a\rVert_{\sharp}-\lVert v_2\rVert_{\sharp}.
 \end{align*}
 Thus $\lVert v_1\rVert_{\sharp}\leq 2\lVert x^{\natural}-a\rVert_{\sharp}+\lVert v_2\rVert_{\sharp}$, which implies that \begin{align}
 \lVert x^{\star}-x^{\natural}\rVert_{\sharp}\leq \lVert v_1\rVert_{\sharp}+\lVert v_2\rVert_{\sharp}\leq 2\lVert x^{\natural}-a\rVert_{\sharp}+2\lVert v_2\rVert_{\sharp}\leq  2\lVert x^{\natural}-a\rVert_{\sharp}+2L\lVert x^{\star}-x^{\natural}\rVert_2. \label{1}
 \end{align}
 Assume that $\lVert x^{\star}-x^{\natural}\rVert_2>C_1\varepsilon$, since otherwise the desired result holds. Then, we have $$
 \lVert \Phi x^{\star}-\Phi x^{\natural}\rVert_{\ell_p}\leq  \lVert \Phi x^{\star}-(\Phi x^{\natural}+e)\rVert_{\ell_p}+\lVert e\rVert_{\ell_p}\leq 2\varepsilon<\frac{2}{C_1}\lVert x^{\star}-x^{\natural}\rVert_2=\alpha\lVert x^{\star}-x^{\natural}\rVert_2
 $$
 by taking $C_1=\frac{2}{\alpha}$. Therefore, by (a), it holds that \begin{align}
 \lVert x^{\star}-x^{\natural}\rVert_2\leq \rho\lVert x^{\star}-x^{\natural}\rVert_{\sharp}.\label{2}
 \end{align}
Therefore, by substituting (\ref{2}) into (\ref{1}), we have $$
\lVert x^{\star}-x^{\natural}\rVert_{\sharp}\leq\frac{2}{1-2\rho L}\lVert x^{\natural}-a\rVert_{\sharp}\leq 4\lVert x^{\natural}-a\rVert_{\sharp}
$$
provided that $\rho\leq\frac{1}{4L}$. Finally, applying (\ref{2}) again implies that $$
 \lVert x^{\star}-x^{\natural}\rVert_2\leq \rho\lVert x^{\star}-x^{\natural}\rVert_{\sharp}\leq 4\rho\lVert x^{\natural}-a\rVert_{\sharp}=C_0\lVert x^{\natural}-a\rVert_{\sharp}\leq C_0\lVert x^{\natural}-a\rVert_{\sharp}+C_1\varepsilon
$$
with $C_0=4\rho$.\\

 $(b)\Rightarrow(a)$: Pick $x^{\natural}$ such that $\lVert \Phi x^{\natural}\rVert_{\ell_p}<\alpha\lVert x^{\natural}\rVert_2$. Then by setting $\varepsilon=\alpha\lVert x^{\natural}\rVert_2$ and $e=0$, we observe that $x^{\star}=0$ is a feasible solution of (\ref{min}), and so with $a=0$ we obtain $$
 \lVert x^{\natural}\rVert_2=\lVert x^{\star}-x^{\natural}\rVert_2\leq C_0\lVert x^{\natural}\rVert_{\sharp}+C_1\varepsilon=C_0\lVert x^{\natural}\rVert_{\sharp}+\alpha C_1\lVert x^{\natural}\rVert_2.
 $$
 Thus, we have $$
 \lVert x^{\natural}\rVert_2\leq \frac{C_0}{1-\alpha C_1} \lVert x^{\natural}\rVert_{\sharp}=\rho\lVert x^{\natural}\rVert_{\sharp}
 $$
 by taking $\alpha=\frac{1}{2C_1}$ and $\rho=2C_0$. 
\section{Robust NSP implies RWP}
In this section, we propose a general version of robust NSP for CS spaces $(\mathcal{H},\mathcal{A},\lVert\cdot\rVert_{\sharp})$ and establish its relationship with RWP.
\begin{definition}
For CS space $(\mathcal{H},\mathcal{A},\lVert\cdot\rVert_{\sharp})$ with bounds $L$ and $p\geq 1$, we say a linear operator $\Phi:\mathcal{H}\rightarrow \mathbb{F}^m$ satisfies the robust NSP with respect to $\ell_p$ with constants $\phi>0$ and $\tau>0$ if for all $v\in\mathcal{H}$ and $a\in\mathcal{A}$ \begin{align}
\lVert v\rVert_2\leq \phi\lVert v-a\rVert_{\sharp}+\tau\lVert \Phi v\rVert_{\ell_p}
\end{align}	
\end{definition}

In fact, this kind of definition of robust NSP is a weaker version compared to the traditional ones we used before. Several examples are list here. \\

\noindent
{\bf Example 1} Take $\mathcal{H}$ to be $\mathbb{C}^N$, $\mathcal{A}=\Sigma_s:=\{x:\lVert x\rVert_{\ell_0}:=\mathrm{card}(\{j\in[N]:x_j\neq 0\})\leq s\}$ and $\lVert\cdot\rVert_{\sharp}=\lVert\cdot\rVert_{\ell_1}$. Then the traditional robust NSP with respect to $\ell_p$ norm of order $s$ with constants $0<\psi<1$ and $\tau>0$ \cite{dlr,fr} is defined as for any set $S\subset [N]$ with $|S|\leq s$ and any $v\in \mathbb{C}^N$, $$
\lVert v_S\rVert_{\ell_2}\leq \frac{\psi}{\sqrt{s}}\lVert v_{S^c}\rVert_{\ell_1}+\tau\lVert \Phi v\rVert_{\ell_p}.
$$
If we take $S$ to be the index set of $s$ largest (in modulus) entries of $v$, then for any $v\in \mathbb{C}^N$ we have \begin{align*}
\lVert v\rVert_{\ell_2}&\leq \lVert v_S\rVert_{\ell_2}+\lVert v_{S^c}\rVert_{\ell_2}\\
&\leq \lVert v_S\rVert_{\ell_2}+\lVert v_{S^c}\rVert_{\ell_1}\\
&\leq \frac{\psi}{\sqrt{s}}\lVert v_{S^c}\rVert_{\ell_1}+\tau\lVert \Phi v\rVert_{\ell_p}+\lVert v_{S^c}\rVert_{\ell_1}\\
&=(\frac{\psi}{\sqrt{s}}+1)\lVert v_{S^c}\rVert_{\ell_1}+\tau\lVert \Phi v\rVert_{\ell_p}\\
&\leq (\frac{\psi}{\sqrt{s}}+1)\lVert v-a\rVert_{\ell_1}+\tau\lVert \Phi v\rVert_{\ell_p}.
\end{align*}
Therefore, the robust NSP given in Definition 3 holds with $\phi= (\frac{\psi}{\sqrt{s}}+1)$. Similar conclusion holds in the weighted sparsity case. Specifically, for every $v\in\mathbb{C}^N$ and $0<q\leq 2$, we define $\lVert v\rVert_{w,q}=\left(\sum\limits_{j=1}^N w_j^{2-q}|v_j|^q\right)^{1/q}$ with $w=(w_j)_{j\in[N]}$ being a vector of weights $w_j\geq 1$. Set $\lVert v\rVert_{\sharp}=\lVert v\rVert_{w,1}$ and $\mathcal{A}=\Sigma_{w,s}:=\left\{x:\sum\limits_{\{j:x_j\neq 0\}}w_j^2\leq s\right\}$. Then the weighted robust NSP with respective to $\ell_p$ norm of order $s$ with constants $\psi\in(0,1)$ and $\tau>0$ (the case $p=2$ see the Definition 4.1 in \cite{rw}): $$
\lVert v_S\rVert_{\ell_2}\leq \frac{\psi}{\sqrt{s}}\lVert v_{S^c}\rVert_{w,1}+\tau\lVert \Phi v\rVert_{\ell_p},\,\,\,\text{for all $v\in\mathbb{C}^N$ and any set $S\subset [N]$ with $w(S):=\sum\limits_{j\in S}w_j^2\leq s$}
$$
will also imply the robust NSP given in Definition 3.\\

\noindent
{\bf Example 2} Take $\mathcal{H}$ to be the Hilbert space of $C^{n_1\times n_2}$ with inner product $\langle X,Y\rangle=Tr(XY^{*})$, and consider the nuclear norm $\lVert X\rVert_{\sharp}=\lVert X\rVert_{*}$. Let $\mathcal{A}$ denote the sets of matrices with rank at most $r$. With the Schatten $q$-norm of $X\in C^{n_1\times n_2}$ given by $\lVert X\rVert_p=\left(\sum\limits_{j=1}^{n}\sigma_j(X)^q\right)^{1/q} (q\geq 1)$ where $\sigma_j(X), j=1,\cdots,n$ denote the singular values of $X$, then the nuclear norm $\lVert \cdot\rVert_{*}$ and the Frobenius norm $\lVert \cdot\rVert_F$ correspond to the cases that $q=1$ and $q=2$, respectively. We write $X=X_r+X_c$, where $X_r$ is the best rank $r$ approximation of $X$ with respect to any Schatten $q$-norm of $X$. Then, for $p\geq 1$, we say that $\Phi:\mathbb{C}^{n_1\times n_2}\rightarrow \mathbb{C}^m$ satisfies the Frobenius-robust rank NSP with respect to $\ell_p$ of order $r$ with constants $0<\psi<1$ and $\tau>0$ \cite{kkrt} if for all $M\in\mathbb{C}^{n_1\times n_2}$, the singular values of $M$ satisfy 
$$
\lVert M_r\rVert_{2}\leq \frac{\psi}{\sqrt{r}}\lVert M_c\rVert_{1}+\tau\lVert \Phi M\rVert_{\ell_p}.
$$
Then, for all $M$, it is obvious that \begin{align*}
\lVert M_r\rVert_{2}&\leq\lVert M_r\rVert_{2}+\lVert M_c\rVert_{2} \\
&\leq\lVert M_r\rVert_{2}+\lVert M_c\rVert_{1} \\
&\leq \frac{\psi}{\sqrt{r}}\lVert M_c\rVert_{1}+\tau\lVert \Phi M\rVert_{\ell_p}+\lVert M_c\rVert_{1}\\
&\leq  (\frac{\psi}{\sqrt{r}}+1)\lVert M_c\rVert_{1}+\tau\lVert \Phi M\rVert_{\ell_p}\\
&\leq (\frac{\psi}{\sqrt{r}}+1)\lVert M-a\rVert_{1}+\tau\lVert \Phi M\rVert_{\ell_p}.
\end{align*}
holds for all $a\in\mathcal{A}$. Therefore, the robust NSP given in Definition 3 holds in this specific CS space. \\

Next, we show that compared to the robust NSP given above, RWP is actually a weaker property. 

\begin{theorem}
	For any CS space $(\mathcal{H},\mathcal{A},\lVert\cdot\rVert_{\sharp})$ with bound $L$, if a linear operator $\Phi:\mathcal{H}\rightarrow \mathbb{F}^m$	
	satisfies the robust NSP with respect to $\ell_p$ with constants $\phi>0$ and $\tau>0$, then this linear operator $\Phi$ satisfies the $(p,\rho,\alpha)$-RWP over $B_{\sharp}$ with $\rho=2\phi$ and $\alpha=\frac{1}{2\tau}$.
\end{theorem}
\noindent
{\bf Proof.} If for any $x\in\mathcal{H}$ such that $\lVert\Phi x\rVert_{\ell_p}<\alpha\lVert x\rVert_2$, then by applying the robust NSP with respect to $\ell_p$ of $\Phi$, we have \begin{align*}
\lVert x\rVert_2\leq \phi\lVert x-a\rVert_{\sharp}+\tau\lVert \Phi x\rVert_{\ell_p}\leq \phi\lVert x-a\rVert_{\sharp}+\alpha\tau\lVert x\rVert_{2}
\end{align*}
holds for all $a\in\mathcal{A}$. Let $a=0$ and rearrange the inequality, we obtain that $$
\lVert x\rVert_2\leq \frac{\phi}{1-\alpha\tau}\lVert x\rVert_{\sharp}.
$$
Thus, $\lVert x\rVert_2\leq\rho\lVert x\rVert_{\sharp}$ holds by assuming that $\rho=2\phi$ and $\alpha=\frac{1}{2\tau}$.

\section{A direct proof of $RIP_{p,2}$ implies RWP}
Basically, it's often inconvenient to show the robust NSP directly for a given matrix. Thus, one usually works instead with a bit stronger properties, i.e., various versions of RIP. And it's well known that RIP implies part (b) in Theorem 1. Then, one could come into conclusion that RIP implies the RWP based on the equivalence of RWP to the stable and robust recovery result. For the sake of completeness, we will give a direct proof of RIP implies RWP in this section. In addition, we merely focus on the traditional sparsity case for CS space here instead of different versions of RIP for different CS spaces. The similar proofs hold for other cases. To handle the $\ell_p$-constrained CS, the following $RIP_{p,2}$ \cite{jhf} is used.
 
\begin{definition}
	We say $\Phi$ satisfies the $(\mu_{p,2},s,\delta)$-$RIP_{p,2}$ if \begin{align}
	\mu_{p,2}(1-\delta)\lVert x\rVert_{\ell_2}\leq \lVert \Phi x\rVert_{\ell_p}\leq \mu_{p,2}(1+\delta)\lVert x\rVert_{\ell_2}
	\end{align}
	for every $s$-sparse vector $x$. 
\end{definition}

It is known that the stable and robust recovery result is guaranteed if $\Phi$ satisfies the $(\mu_{p,2},s,\delta)$-$RIP_{p,2}$ with some small $\delta$. But this RIP condition is strictly stronger than the RWP condition. Next, we are going to present the fact and proof of that the RIP with a small constant $\delta$ does imply the RWP. Before that, the following lemma which will be used in this proof is provided. It is a direct extension of Lemma 12 in \cite{cm} from $p=2$ to $p\geq 1$. 
\begin{lemma}
Suppose that $\Phi$ satisfies the right-hand inequality of  $(\mu_{p,2},s,\delta)$-$RIP_{p,2}$. Then for every $x$ such that $\lVert x\rVert_{\ell_2}>\rho\lVert x\rVert_{\ell_1}$, we have \begin{align*}
\lVert x-x_{S}\rVert_{\ell_2}<\frac{1}{\rho\sqrt{s}}\lVert x\rVert_{\ell_2},\,\,\,\lVert \Phi(x-x_S)\rVert_{\ell_p}<\frac{(1+\delta)\mu_{p,2}}{\rho\sqrt{s}}\lVert x\rVert_{\ell_2},
\end{align*}
where $S$ denotes the indices of the $s$ largest (in modulus) entries of $x$.
\end{lemma}

\noindent
{\bf Proof.} Let $S_0=S$, and for $j\geq 1$, $S_j$ denote the indices of the $s$ largest (in modulus) entries of $x$ not covered by $S_i$ for $i<j$. Then, for all $j\geq 0$, it holds that $$
\lVert x_{S_{j+1}}\rVert_{\ell_2}\leq \sqrt{s}\max\limits_{i\in S_{j+1}}|x_i|\leq \sqrt{s}\min\limits_{i\in S_{j}}|x_i|\leq \frac{1}{\sqrt{s}}\lVert x_{S_j}\rVert_{\ell_1}.
$$
Therefore, for every $x$ such that $\lVert x\rVert_{\ell_2}>\rho\lVert x\rVert_{\ell_1}$, $$
\lVert x-x_{S}\rVert_{\ell_2}\leq\sum\limits_{j\geq 1}\lVert x_{S_j}\rVert_{\ell_2}\leq \frac{1}{\sqrt{s}}\sum\limits_{j\geq 0}\lVert x_{S_j}\rVert_{\ell_1}=\frac{1}{\sqrt{s}}\lVert x\rVert_{\ell_1}<\frac{1}{\rho\sqrt{s}}\lVert x\rVert_{\ell_2}.
$$
Similarly, it holds that $$
\lVert \Phi(x-x_S)\rVert_{\ell_p}\leq \sum\limits_{j\geq 1}\lVert \Phi x_{S_j}\rVert_{\ell_p}\leq (1+\delta)\mu_{p,2}\sum\limits_{j\geq 1}\lVert x_{S_j}\rVert_{\ell_2}<\frac{(1+\delta)\mu_{p,2}}{\rho\sqrt{s}}\lVert x\rVert_{\ell_2}.
$$
\begin{theorem}
	Suppose $\Phi$ satisfies $(\mu_{p,2},s,\delta)$-$RIP_{p,2}$ with $\delta<\frac{1}{3}$. Then $\Phi$ satisfies the $(p,\rho,\alpha)$-RWP over $B_{\ell_1}$ with \begin{align*}
	\rho=\frac{3}{\sqrt{s}},\,\,\,\alpha=(\frac{1}{3}-\delta)\mu_{p,2}.
	\end{align*}
\end{theorem}

\noindent
{\bf Proof.} The proof procedure follows from the proof of theorem 11 in \cite{cm} with minor modification. It will be proved by contradiction. We assume $\Phi$ satisfies the right-hand inequality of $(\mu_{p,2},s,\delta)$-$RIP_{p,2}$, but violates the $(p,\rho,\alpha)$-RWP, then the proof of the theorem is completed if we can show that $\Phi$ necessarily violates the left-hand inequality of $(\mu_{p,2},s,\delta)$-$RIP_{p,2}$.
 
In fact, if we pick $x$ such that $\lVert \Phi x\rVert_{\ell_p}<\alpha\lVert x\rVert_{\ell_2}$ but $\lVert x\rVert_{\ell_2}>\rho\lVert x\rVert_{\ell_1}$, then we have \begin{align}
\lVert \Phi x_S\rVert_{\ell_p}\leq \lVert \Phi x\rVert_{\ell_p}+\lVert \Phi (x-x_S)\rVert_{\ell_p}<\alpha\lVert x\rVert_{\ell_2}+\frac{(1+\delta)\mu_{p,2}}{\rho\sqrt{s}}\lVert x\rVert_{\ell_2}=\frac{2}{3}(1-\delta)\mu_{p,2}\lVert x\rVert_{\ell_2}
\end{align}
by using Lemma 1 and our specific choices of $\rho$ and $\alpha$. By adopting Lemma 1 again, it holds that $$
\lVert x\rVert_{\ell_2}\leq \lVert x_S\rVert_{\ell_2}+\lVert x-x_S\rVert_{\ell_2}<\lVert x_S\rVert_{\ell_2}+\frac{1}{3}\lVert x\rVert_{\ell_2},
$$
which implies $\lVert x\rVert_{\ell_2}<\frac{3}{2}\lVert x_S\rVert_{\ell_2}$. Therefore, we have $\lVert \Phi x_S\rVert_{\ell_p}<\frac{2}{3}(1-\delta)\mu_{p,2}\lVert x\rVert_{\ell_2}=(1-\delta)\mu_{p,2}\lVert x_S\rVert_{\ell_2}$. This is the desired result and completes the proof.

\section{Recovery results via RWP}

In this section, we will provide the recovery results for $\ell_p$-constrained CS directly via RWP and find the required minimum number of measurements for standard Gaussian  matrices. We take $\mathcal{H}=\mathbb{R}^N$ and $\mathbb{F}=\mathbb{R}$ at the remainder of this section. To obtain the main result, it will be much more convenient to adopt the contrapositive statement of $(p,\rho,\alpha)$-RWP: for any $p\geq 1$, a linear operator $\Phi:\mathbb{R}^N\rightarrow \mathbb{R}^m$ satisfies the $(p,\rho,\alpha)$-RWP over $B_{\ell_1}$ if and only if $\lVert \Phi x\rVert_{\ell_p}\geq \alpha\lVert x\rVert_{\ell_2}$ for every $x\in\mathbb{R}^N$ such that $\lVert x\rVert_{\ell_2}>\rho\lVert x\rVert_{\ell_1}$. Then, by scaling, $(p,\rho,\alpha)$-RWP is further equivalent to having $\lVert \Phi x\rVert_{\ell_p}\geq \alpha$ for every $x\in\mathbb{R}^N$ such that $\lVert x\rVert_{\ell_2}=1$ and $\lVert x\rVert_{\ell_1}<\rho^{-1}$. Taking $T=\rho^{-1}B_{\ell_1}\cap \mathbb{S}^{N-1}$ where $\mathbb{S}^{N-1}$ denotes the unit sphere in $\mathbb{R}^N$, we have the following lemma: 

\begin{lemma}
A linear operator $\Phi:\mathbb{R}^N\rightarrow \mathbb{R}^m$ satisfies $(p,\rho,\alpha)$-RWP over $B_{\ell_1}$ if and only if $\lVert \Phi x\rVert_{\ell_p}\geq \alpha$ for every $x\in T$.
\end{lemma}

Let $X_1,\cdots,X_m$ be i.i.d. copies of a random vector $X$ in $\mathbb{R}^N$. Then we have the following key lemma concerning a lower bound for empirical processes:

\begin{lemma}(Lemma III.1. in \cite{dlr})
	Fix $1\leq p<\infty$, let $\mathcal{F}$ be a class of function from $\mathbb{R}^N$ into $\mathbb{R}$. Consider $Q_{\mathcal{F}}(u)=\inf\limits_{f\in\mathcal{F}}P(|f(X)|\geq u)$ and $R_{m}(\mathcal{F})=E\sup\limits_{f\in\mathcal{F}}|\frac{1}{m}\sum\limits_{i=1}^m\varepsilon_i f(X_i)|$, where $\{\varepsilon_i\}_{i\geq 1}$ is a Rademacher sequence. Let $u>0$ and $t>0$, then, with probability at least $1-2e^{-2t^2}$, $$
	\inf\limits_{f\in\mathcal{F}}\frac{1}{m}\sum\limits_{i=1}^m|f(X_i)|^p\geq u^p\left(Q_{\mathcal{F}}(2u)-\frac{4}{u}R_m(\mathcal{F})-\frac{t}{\sqrt{m}}\right)
	$$
\end{lemma}

With the above two lemmas being applied, the main recovery result for standard Gaussian matrices is provided as follows:

\begin{theorem}
	 Take $\rho>0$  and Let $\Phi$ be an $m\times N$ standard Gaussian matrix. Define the Gaussian width of $T$ as $w(T)=E \sup_{x\in T}\langle g,x\rangle$, where $g$ has i.i.d. $N(0,1)$ entries. For any $p\geq 1$, set $\alpha=c_1m^{1/p}$ and if $$
	m\geq c_0 w(T)^2,
	$$
	then $\Phi$ satisfies the $(p,\rho,\alpha)$-RWP over $B_{\ell_1}$ with probability exceeding $1-2e^{-c_2 m}$.
\end{theorem}

\noindent
{\bf Proof.}  According to Lemma 2, it suffices to show that $$
P\left(\inf\limits_{x\in T}\lVert \Phi x\rVert_{\ell_p}\geq \alpha\right)\geq 1-2e^{-c_2 m}.
$$
We write $$
\inf\limits_{x\in T}\lVert \Phi x\rVert_{\ell_p}=\inf\limits_{x\in T}\left(\sum\limits_{i=1}^m|\langle X_i,x\rangle|^p\right)^{1/p},
$$
where $X_i$ denotes the $i$-th row of $\Phi$. 

Firstly, we suppose $1\leq p<\infty$, then applying Lemma 3 with $\mathcal{F}=\{\langle \cdot,x\rangle: x\in T\}$, we obtain that \begin{align}
\inf\limits_{x\in T}\lVert \Phi x\rVert_{\ell_p}=\inf\limits_{x\in T}\left(\sum\limits_{i=1}^m|\langle X_i,x\rangle|^p\right)^{1/p}\geq m^{1/p}u\left(Q_{\mathcal{F}}(2u)-\frac{4}{u}R_m(\mathcal{F})-\frac{t}{\sqrt{m}}\right)^{1/p}.
\end{align}
holds with probability at least $1-2e^{-2t^2}$. Next, we estimate the small ball probability $Q_{\mathcal{F}}(2u)$ and the expected Rademacher supremum $R_{m}(\mathcal{F})$ respectively. Since for any $x\in \mathbb{S}^{N-1}$, $\lVert x\rVert_{\ell_2}=1$, therefore $$
P(|\langle X_i,x\rangle|\geq u )=P(|G|\geq u),
$$
where $G$ is a standard Gaussian real-valued random variable. Hence, it holds that \begin{align}
Q_{\mathcal{F}}(2u)\geq P(|G|\geq 2u). \label{q}
\end{align}
In addition, let $V=\frac{1}{\sqrt{m}}\sum\limits_{i=1}^m\varepsilon_i X_i$, then $V$ is a standard Gaussian vector. Then we have \begin{align*}
R_{m}(\mathcal{F})=\frac{1}{\sqrt{m}}E\sup\limits_{x\in T}\langle V,x\rangle=\frac{1}{\sqrt{m}}w(T).
\end{align*}
Now pick $u_{\star}$ small enough such that the right side of (\ref{q}) is bigger than $\frac{1}{2}$ and $t=\sqrt{\frac{c_2 m}{2}}$ with $c_2=2\left(\frac{1}{2}-\frac{4}{u_{\star}\sqrt{c_0}}-(\frac{c_1}{u_{\star}})^p\right)^2$, then with probability exceeding $1-2e^{-c_2m}$ it holds that \begin{align*}
\inf\limits_{x\in T}\lVert \Phi x\rVert_{\ell_p}&\geq m^{1/p}u_{\star}\left(\frac{1}{2}-\frac{4}{u_{\star}}\frac{1}{\sqrt{m}}w(T)-\frac{t}{\sqrt{m}}\right)^{1/p} \\
&\geq m^{1/p}u_{\star}\left(\frac{1}{2}-\frac{4}{u_{\star}\sqrt{c_0}}-\sqrt{\frac{c_2}{2}}\right)^{1/p} \\
&=: c_1 m^{1/p}=\alpha.
\end{align*}

Finally, when $p=\infty$, since $\lVert \Phi x\rVert_{\ell_{\log m}}\leq e\lVert \Phi x\rVert_{\ell_\infty}$, hence \begin{align*}
P(\inf\limits_{x\in T}\lVert \Phi x\rVert_{\ell_\infty}\geq c_1)\geq P(\inf\limits_{x\in T}\lVert \Phi x\rVert_{\ell_{\log m}}\geq c_1 e).
\end{align*}
Thus, the result follows from the already proved case that $p=\log m$ and the proof is completed.\\

To further compute the lower bound of the required number of measurements $m$, we need the following upper bound for the Gaussian width of set $T$:

\begin{lemma}(Lemma 16 in \cite{cm})
	There exists an absolute constant c such that \begin{align}
	w(\sqrt{s}B_{\ell_1}\cap \mathbb{S}^{N-1})\leq c\sqrt{s\log(eN/s)}
	\end{align}
	for every positive integer $s$.
\end{lemma}

Then, the order of the minimum required number of measurements $m$ for standard Gaussian matrices to guarantee the stable and robust recovery result with high probability is established here. A similar result is obtained in Theorem III.3 of \cite{dlr} via robust NSP. Since it's shown in Section 3 that RWP is weaker than robust NSP, thus we can weaken the condition of the required minimum $m$ up to constants although it seems that the obtained orders are the same and optimal via both methods. 

\begin{cor}
	Let $\Phi$ be an $m\times N$ standard Gaussian matrix. Fix $1\leq p\leq \infty$ and $0<\eta<1$. There exists some constants $\{C, D, C', D'\}$ such that if $$
	m\geq Cs\log(eN/s)+D\log(2\eta^{-1}),
	$$
then with probability exceeding $1-\eta$, for every $x^{\natural}\in\mathbb{R}^N$, $y=\Phi x^{\natural}+e$ with $e\in\mathbb{R}^m$ and $\lVert e\rVert_{\ell_p}\leq \varepsilon$, any solution $x^{\star}$ to  \begin{align}
\min\limits_{x}\,\,\lVert x\rVert_{\ell_1}\,\,\,\text{subject to}\,\,\,\lVert\Phi x-y\rVert_{\ell_p}\leq \varepsilon
\end{align} 
satisfies $\lVert x^{\star}-x^{\natural}\rVert_{\ell_2}\leq \frac{C'}{\sqrt{s}}\sigma_s(x^{\natural})_1+\frac{D'}{m^{1/p}}\varepsilon$, where $\sigma_s(x^{\natural})_1=\inf\limits_{\lVert z\rVert_{\ell_0}\leq s}\lVert x^{\natural}-z\rVert_{\ell_1}$.
\end{cor} 

\noindent
{\bf Proof.} Take the CS space $(\mathcal{H},\mathcal{A},\lVert\cdot\rVert_{\sharp})$ with bound $L$ to be the traditional sparse space with $\mathcal{H}=\mathbb{R}^N$, $\mathcal{A}=\Sigma_s$ and $\lVert\cdot\rVert_{\sharp}=\lVert\cdot\rVert_{\ell_1}$. Then as given in \cite{cm}, we have $L=\frac{1}{\sqrt{s}}$. Set $\rho=\frac{C'}{4\sqrt{s}}\leq \frac{1}{4\sqrt{s}}$ with $0<C'\leq 1$ and $\alpha=\frac{2}{D'}m^{1/p}$, then the desired result follows from Theorem 1 if the $(p,\rho,\alpha)$-RWP over $B_{\ell_1}$ of $\Phi$ is verified. In fact, by adopting Theorem 4 and Lemma 4, this holds when $m\geq Cs\log(eN/s)+D\log(2\eta^{-1})$ with $C$ and $D$ being appropriately chosen.

\section{Conclusion}
In this paper, we studied the equivalent RWP condition to guarantee the stable and robust recovery result for $\ell_p$-constrained CS procedure. We proposed a general version of robust NSP and verified that both robust NSP and RIP imply RWP. Finally, the specific recovery condition for standard Gaussian matrices was established via RWP and the optimal order of the minimum required number of measurements was obtained. The present paper extends the CS recovery analysis results via RWP in \cite{cm} from $\ell_2$-constrained to $\ell_p$-constrained case with $p\geq 1$, and also largely complements the existing theoretical results for robust CS with $\ell_p$ loss function, see \cite{agr,dlr,jhf,kkrt,sac,wllqy}, and among others.

\end{document}